# Phonon-Mediated Attractive Interactions between Excitons in Lead-Halide-Perovskites

Nuri Yazdani[1,2,3*], Maryna I. Bodnarchuk[4,5], Federica Bertolotti[6], Norberto Masciocchi[6], Ina Fureraj[7], Burak Guzelturk[8], Benjamin L. Cotts[1,9], Marc Zajac[8], Gabriele Rainò,[4,5] Maximilian Jansen,[3] Simon C. Boehme,[4,5] Maksym Yarema[10], Ming-Fu Lin[11], Michael Kozina[11], Alexander Reid[11], Xiaozhe Shen[11], Stephen Weathersby[11], Xijie Wang[11], Eric Vauthey[7], Antonietta Guagliardi[6,12], Maksym V. Kovalenko[4,5], Vanessa Wood[3*], Aaron Lindenberg[1,2,13,14*]

1. Department of Materials Science and Engineering, Stanford University, Stanford, CA, USA.
2. Stanford Institute for Materials and Energy Sciences, SLAC National Accelerator Laboratory, Menlo Park, CA, USA.
3. Department of Information Technology and Electrical Engineering, ETH Zürich, Zürich, Switzerland.
4. Department of Chemistry and Applied Biosciences, ETH Zürich, Zürich CH-8093, Switzerland.
5. Empa- Swiss Federal Laboratories for Materials Science and Technology, Dübendorf, Dübendorf CH-8600, Switzerland
6. Dipartimento di Scienza e Alta Tecnologia & To.Sca.Lab, Università dell'Insubria, 22100 Como, Italy.
7. Department of Physical Chemistry, University of Geneva, 30 Quai Ernest-Ansermet, CH-1211 Geneva 4, Switzerland.
8. X-ray Science Division, Argonne National Laboratory, Lemont, Illinois 60439, United States.
9. Department of Chemistry and Biochemistry, Middlebury College, Middlebury, VT, USA.
10. Chemistry and Materials Design Group, Institute for Electronics, Department of Information Technology and Electrical Engineering, ETH Zurich, Gloriastrasse 35, 8092 Zurich, Switzerland.
11. Linac Coherent Light Source, SLAC National Accelerator Laboratory, Menlo Park, CA, USA.
12. Istituto di Cristallografia & To.Sca.Lab, Consiglio Nazionale delle Ricerche, 22100 Como, Italy.
13. The PULSE Institute for Ultrafast Energy Science, SLAC National Accelerator Laboratory, Menlo Park, CA, USA.
14. Department of Photon Science, Stanford University and SLAC National Accelerator Laboratory, Menlo Park, CA, USA.

* nuri@ethz.ch, vwood@ethz.ch, aaronl@stanford.edu

## Abstract

Understanding the origin of electron-phonon coupling in lead-halide perovskites (LHP) is key to interpreting and leveraging their optical and electronic properties. Here we perform femtosecond-resolved, optical-pump, electron-diffraction-probe measurements to quantify the lattice reorganization occurring as a result of photoexcitation in LHP nanocrystals. Photoexcitation is found to drive a reduction in lead-halide octahedra tilts and distortions in the lattice, a result of deformation potential coupling to low energy optical phonons. Our results indicate particularly strong coupling in $FAPbBr_3$, and far weaker coupling in $CsPbBr_3$, highlighting differences in the dominant mechanisms governing electron-phonon coupling in LHPs. We attribute the enhanced coupling in $FAPbBr_3$ to its disordered crystal structure, which persists down to cryogenic temperatures. We find the reorganizations induced by each exciton in a multiexcitonic state constructively interfere, giving rise to a coupling strength which scales quadratically with the exciton number. This superlinear scaling induces phonon-mediated attractive interactions between excitations in LHPs.

**Main Text**

Lead-halide perovskite (LHP) have advanced to the forefront of materials research for a wide array of applications including optoelectronic devices (e.g., solar cells),[1,2] near unity quantum yield light sources,[3] and coherent single-photon emitters for quantum information processing.[4] Electron-phonon coupling (EP-coupling) plays a critical role in LHPs, expected to both enhance performance metrics in some cases (e.g. 'polaronic protection' of charge carriers)[5] and limit them in others (e.g. exciton coherence loss and broadened emission in perovskite nanocrystals).[6]

While there has been extensive discussion of the coupling to the highest energy longitudinal optical (LO) phonon (e.g., ~17 meV in lead-bromides) to interband transitions in these systems,[7–10] there is a growing appreciation for the importance of the lower energy optical modes (e.g., ~2.5 - 12 meV in lead-bromides),[11–13] particularly in the hybrid lead-halides where their coupling can outweigh that to the high energy LO mode.[14–18] At the root of EP-coupling is a shift of the equilibrium atomic coordinates of the atoms in a material upon a change of the electronic configuration (**Fig. 1a**).[19] While various time-resolved spectroscopies have shed light on the time-scales of the photoexcitation-induced lattice reorganization and the phonons involved,[9,11,20,21] the nature of the reorganization, and therefore the mechanisms underlying the coupling, remain unclear. Valuable insight can be provided through physical characterization of the inherent excited-states structural dynamics of these materials.[21–23] In principle, lattice reorganization can be directly measured through time-resolved diffraction. In a semiconductor nanocrystal (NC), the size of which is comparable or smaller than the exciton radius, such reorganization is expected to occur over the entire volume of the NC. Furthermore, NCs offer the possibility of exciting large numbers of excitons simultaneously withinin the same NC volume, which can enhance the magnitude of the lattice reorganization facilitating its detection.

Here, we perform time-resolved, optical-pump, electron-diffraction-probe measurements to quantify the lattice reorganization occurring as a result of EP-coupling to the interband transition in formamidinium-lead-bromide (FAPbBr$_3$, FA = CH$_5$N$_2$) NCs. We observe that excitons drive a reorganization of the Pb-Br sublattice towards higher symmetry, which is in contrast to the Frohlich polaron picture in which one expects a decrease in overall symmetry as a result of lattice polarization. To explain this surprising finding, we develop a deformation-potential EP-coupling model based on the fact that in LHPs the reduction of octahedral tilting and distortions drive a red-shift renormalization of the bandgap,[24–27] which reduces the total energy of the excitons. We then use our model to extract EP-coupling strengths (Huang-Rhys factors) directly from the time-resolved measurements. Our findings provide an intuitive explanation for the origin of low energy optical phonon coupling in LHPs, and link the strong coupling to these modes in FAPbBr$_3$ NCs to its locally tilted/disorderd crystal structure,[28,29] which is found to persist down to cryogenic temperatures. Finally, the magnitude of the coupling strength related to octahedral tilting and distortions is found to scale quadratically with the exciton number, inducing a phonon-mediated attractive interaction between excitons.

We perform measurements at the mega-electronvolt ultrafast electron diffraction facility (MeV-UED) at SLAC (**Fig. 1b**) on ~9.5 nm FAPbBr$_3$ and CsPbBr$_3$ NCs,[30] the size of which are comparable to estimated polaron (6 - 14 nm)[20,21,31] and exciton (7 nm)[32] diameters in lead-bromide perovskites. Measurements are performed over a temperature range of 100 – 280 K with pump fluences of 0.07-0.8 mJ/cm$^2$. 400 nm pump photons are ~650 meV above the bandgap of the NCs and generate exciton densities $N_{ex}$ ~ 5 - 50 excitons per NC.[33] Experimental details are found in the **Methods**. From the measured time resolved diffraction, $I(t,q)$, we compute the differential scattering intensity as a function of time $t$ and momentum transfer $q$,

$$\Delta I(t,q) = (I(t,q) - I_0(q))/I_0(q), (1)$$

where $I_0(q)$ is the measured scattering of the sample in the absence of photoexcitation. The plot of $\Delta I(t,q)$ at 100 K with 0.8 mJ/cm$^2$ is shown in **Fig. 1c** and reveals a fully reversible reorganization of the FAPbBr$_3$ lattice upon photoexcitation, with large, fast changes in diffraction intensities at specific momentum transfers, $q$. Under the same experimental conditions, no lattice response in the CsPbBr$_3$ NCs is discernable within the statistics of the measurement (**Fig. 1d**).

The time-scales of the dynamics of $\Delta I(t,q)$ are independent of $q$, as demonstrated in **Fig. 2a**, where $\Delta I(t,q)$ is plotted at specific $q$ values. To quantify the dynamics, we fit the differential scattering with a biexponential function, $\sim \exp[-t/\tau_S] - \exp[-t/\tau_L]$, from which we extract the timescale for the onset of the lattice reorganization upon excitation, $\tau_S$, and for the return of the lattice to equilibrium, $\tau_L$ (**Supplementary Note 1, Fig. S2**). The onset occurs on a timescale of $\tau_S \sim 1.4$ ps, irrespective of the pump-fluence and temperature indicating a time-scale intrinsic to the structural response of FAPbBr$_3$ (see **Table S1**). We note that this timescale is similar to the frequency of low energy optical phonons in the lead-bromide perovskites (0.6 THz).[34] The lattice relaxes back to equilibrium on time scales of $\tau_L \sim 30$ to 50 ps, again with little discernable impact of pump-fluence and temperature, which is within the range of measured multi-exciton decay rates in LHP NCs under similar excitation conditions.[35,36]

We rule out transient heating of the NCs[23] as a cause of the observed lattice response because the measured timescales and magnitude of the lattice reorganization are not consistent with the trends expected due to thermal effects (**Supplementary Note 2**). The time-scales rather point to a picture of lattice reorganization associated with the coupling of the lattice to the interband excitation of excitons. While the dominant EP-coupling in FAPbBr$_3$ has been shown to be to lower energy optical modes,[11,15,16] coupling to the high energy LO phonon dominates in CsPbBr$_3$.[12] The observed strong reorganization of FAPbBr$_3$ and lack of signal in CsPbBr$_3$ therefore suggests that the observed lattice reorganization may potentially be attributed to the coupling of the interband transition to lower energy optical phonons. We find that the magnitude of the differential scattering, $\Delta I(t,q)$, scales linearly with pump-fluence (**Fig. 2b**) as highlighted in **Fig. 2c** for $t = 5$ ps and $q = 2.6$ Å$^{-1}$. This finding indicates that the magnitude of the lattice reorganization is linearly dependent on the exciton number, $N_{ex}$, which is also consistent with the fact that the timescale for the lattice to return to equilibrium $\tau_L$ is the same as that for multi-exciton recombination.[35,36]

We next consider the atomistic origins of the observed lattice reorganization from the measured $\Delta I(t,q)$. As shown in **Fig. 2b** for $t = 5$ ps, the differential scattering upon photoexcitation is characterized primarily by strong reductions in the diffraction intensity at $q$ values $q \sim 2.6, \sim 3.6$ and $\sim 5.6$ Å$^{-1}$, with a mild increase in the scattering at most other $q$. The largest differential feature at 2.6 Å$^{-1}$ corresponds to the 211 peak (using cubic *hkl* indices). The magnitude of this peak is highly sensitive to the extent of octahedral tilting in perovskite structures and is minimized in the cubic phase. In **Fig. 2d**, we plot the simulated intensity of the 211 peak as a function of the Pb-Br-Pb tilt angle, where a linear proportionality is evident for a variety of low-symmetry perovskite structures. The reduction of the 211 peak upon photoexcitation therefore indicates that excitons on the NCs drive a reduction of the magnitude of octahedral tilting within the Pb-Br sublattice.

The additional large negative differentials occur at ~ 3.6 and 5.6 Å$^{-1}$ close to the 222 and 520 peak positions, however these peaks are highly insensitive to distortions of the Pb-Br sublattice (**Supplementary Note 3**). On the contrary, the closely located 311 (~3.45 Å$^{-1}$) and 511 (~5.4 Å$^{-1}$) peaks are highly sensitive to correlated Pb off-centering within the Pb-Br octahedra and are minimized when Pb-atoms lie in the center of regular octahedra. Finite Pb-off-centering has been shown to be intrinsic in the equilibrium structure of FAPbBr$_3$,[37] and simulations of the differential scattering assuming a photo-excited reduction of Pb-off-centering provides a qualitatively consistent description of the observed spectra (**Fig. S3**).

Exciton-phonon coupling to the interband excitation of excitons in FAPbBr$_3$ NCs therefore drives a structural lattice reorganization through reduction of the Pb-Br-Pb tilt angles and incipient regularization of the Pb-Br octahedra. This finding is at odds with the simple picture of polar Fröhlich coupling which would decrease lattice symmetry. Rather, we argue that these findings point to a deformation potential type EP-coupling to phonons which drive changes in the Pb-X-Pb bonding angles within the LHP.

It is known that Pb-X octahedral tilts and distortions impact the bandgap of LHPs.[24–26] Both the valence band (VB) and conduction band (CB) derive from *sp*-bonding of the Pb-X sublattice, with Pb-*s* and X-*p* antibonding about the VB-maximum and X-*s* and Pb-*p* antibonding about the CB minimum.[27] Any deviation of the Pb-X-Pb bonding angles from 180° will decrease the *sp* coupling between atomic orbitals in both bands, reducing their bandwidth, thereby increasing the bandgap of the LHP (**Fig. 3a**). In **Fig. 3b**, we plot the renormalization of the bandgap, using CsPbBr$_3$ as a model system, as a function of *Pnma* octahedral tilting and Pb off-centering, both of which significantly blueshift the bandgap with increasing tilting or Pb displacement. While 180° Pb-X-Pb bonds will minimize the bandgap, minimization of the enthalpy of the lattice determines the equilibrium structure, and LHPs frequently adopt lower symmetry perovskite structures, with finite octahedral tilts and distortion,[28,38] because A-X non-covalent bonds reduce the lattice formation energies of the lower symmetry polymorphs relative to the cubic phase.[38–40]

The strong bandgap renormalization occurring with distortions of the Pb-X-Pb bonding angles means that the phonons in a LHP which drive these distortions in the low symmetry polymorphs will couple to interband transitions. Provided these distortions exist in the equilibrium phase, in the excited state, this coupling will drive a reduction in the magnitudes of these distortions, thus minimizing the exciton energy and increasing the lattice symmetry.

To illustrate this, we take a simple model assuming a single phonon with frequency $\omega$ and dimensionless normal coordinate $Q$, driving an octahedral tilt $\theta_{Pb-X-Pb}$ in an LHP (**Fig. 3c**). In the absence of any excitation, the energy of the lattice is given by

$$E_0(Q) = \frac{1}{2}\hbar\omega Q^2, \quad (2)$$

where $Q = 0$ corresponds to the equilibrium phase with some finite tilt and bandgap $E_{g0}$. We assume a first-order linear scaling of the bandgap along $Q$ (i.e., $\partial E_g/\partial \theta_{Pb-X-Pb} \propto \partial E_g/\partial Q$, see **Fig. 3b** and **Fig. S5**). To first order the energy of each exciton scales proportionaly to the bandgap, and write the total energy upon exciting $N_{ex}$ excitons as

$$E_{N_{ex}}(Q) = \frac{1}{2}\hbar\omega Q^2 + N_{ex}\left(E_{g0} + \frac{\partial E_g}{\partial Q}Q\right). \quad (3)$$

This can be minimized to find the shift of the normal coordinate (magnitude of the lattice reorganization) in the excited state,

$$Q_{N_{ex}} = N_{ex}\left(-\frac{\partial E_g}{\partial Q}\right)/\hbar\omega, (4)$$

which scales linearly with the number of excitons as observed in the experiments (**Fig. 2c**). The EP-coupling strength, typically referred to as the Huang-Rhys factor,[41] is then given by

$$\tilde{S}_{N_{ex}\omega} = Q_{N_{ex}}{}^2/2. (5)$$

In **Supplementary Note 4**, we provide a more detailed mathematical model for this deformation-potential coupling that extends beyond the single phonon assumption.

By computing and analyzing the phonon density of states (**Fig. 3d, Supplementary Note 4**), we demonstrate that it is lower energy optical phonons (~2.5 - 12 meV), which couple to interband transitions as a result of Pb-X-Pb bond-angle distortions. For example, in the ideal orthorhombic *Pnma* structure (with regular octahedra), optical modes at ~3, 6, and 7.5 meV drive the tilting in the Pb-Br LHPs, while 6 modes with energies ~10 - 12 meV drive Pb-displacements. We can extract the EP-phonon coupling strengths of these modes to the excitation of a single exciton in the $FAPbBr_3$ NCs from the MeV-UED results using the measured changes in the 211 and 311 peak intensities (**Fig. 3e**, **Supplementary Note 5**). The strongest coupling of $\tilde{S}_{1\omega} \sim 0.3$ to the 6 meV optical mode is in excellent agreement to that reported from low temperature single dot luminescence measurements, where a coupling to a 5 meV mode of $\sim 0.15 - 0.35$ was estimated for similarly sized NCs.[16] Couplings to the same modes calculated for equivalently sized $CsPbBr_3$ NCs are over an order of magnitude weaker (e.g. $\tilde{S}_{1\omega} \sim 0.01$ for the 6 meV mode, see **Fig. S8**) consistent with previous studies.[7,12] The strong coupling strength in FA and weak coupling in Cs explains why we observe a large lattice reorganization in $FAPbBr_3$ and not in $CsPbBr_3$ (**Fig. 1c,d**). The question remains however as to why the coupling to low energy optical phonons is far stronger in $FAPbBr_3$ relative to $CsPbBr_3$.

To gain insight into the origins of the strong coupling in $FAPbBr_3$ NC, we consider its structure. Bulk $FAPbBr_3$ has been reported to adopt distinct phases for specific temperature regimes i.e, orthorhombic < ~137 K, tetragonal < ~262 K, and cubic > ~262 K.[42] We repeat our time-resolved MeV-UED measurements at temperatures of 100, 200, and 280 K (**Fig. 4a**). Surprisingly, we find similar lattice response at all three temperatures with an increase in the magnitude of the lattice reorganization with increasing temperature (**Fig. 4b**). This finding implies non-zero magnitudes of octahedral tilt reductions and Pb-displacements in the equilibrium structure even at 280 K. This is at odds with the assignment of a simple *Pm3m* cubic perovskite structure with straight Pb-Br-Pb linkages, but is consistent with $FAPbBr_3$ NCs exhibiting disorded Br ions (and locally tilted Pb-Br-Pb angles) within an average-cubic phase.[30] In this case, photoexcitation reduces Pb-Br-Pb bending, pushing the system towards the archetypal *Pm3m* cubic phase. This disordered phase has been described with the 'split-cubic' (SC) perovskite model in which the local Pb-Br-Pb bend angles are finite, but lack any long range order, making the average structure metrically, and structurally, cubic.[43] We note that in the SC structure, the intensity of the 211 peak is highly sensitive to the magnitude of the local Pb-Br-Pb tilts while 311 and 511 peaks are fingerprints of the Pb displacements (**Fig. S4**).

We propose that the observed enhancement in the coupling of low energy optical phonons to interband transitions in $FAPbBr_3$ NCs is linked to its disordered phase. This link would however imply a persistence of disordered structure down to 0 K, as the strong coupling to these modes has been shown to persist at cryogenic temperatures.[15,16] To confirm this, we perform temperature-dependent X-ray total scattering measurements (at the MS-X04SA beamline of the Swiss Light Source) and find no indication of long-range ordering of any low symmetry LHP phase as the temperature is decreased, with the disordered phase observed over the entire measured temperature range (30 to 300 K, **Fig. 4c**). The origin of this disordered phase is likely a result of a glassy state of the FA orientations,[44] with strong correlations between the local octahedral tilts and the actual orientation of the large FA ions of *mm2* symmetry within an "ideal" *m3m* symmetry site.[45,46] In **Supplementary Note 7** we discuss several possible mechanisms which can enhance EP-coupling in the disordered phase and reproduce the observed increase in coupling with temperature, including phonon-softening, sizeable entropic contributions to the free-energy of the FAPbBr$_3$ lattice, and correlations between anharmonic FA reorientation and Pb-Br distortions.[46,47]

Finally, we consider the implications of the strong coupling. For this, we turn our attention back to the finding that the magnitude of the lattice reorganization is linearly dependent on the exciton number, $N_{ex}$ (**Fig. 2c** and **Eq. 4**), which indicates constructive interference of the lattice reorganization. In this case, the EP-coupling strength depends quadratically on the magnitude of the lattice reorganization, this indicates a coupling strength which scales quadratically with the exciton number $\tilde{S}_{N_{ex}\omega} \propto N_{ex}^2$ (**Eq. 5**). This quadratic scaling of $\tilde{S}_{N_{ex}\omega}$ leads to massive reorganization energies, $\lambda_{Nex}$, associated with multiexcitonic states. With the coupling extracted for the FAPbBr$_3$ NCs (**Fig. 3e**), the reorganization energy of a 20 exciton state would be $\lambda_{Nex} \sim \sum_\omega \tilde{S}_{1\omega} N_{ex}^2 \hbar\omega \sim 2.8$ eV.

This can be experimentally corroborated through measurement of the energy of photons emitted from the multiexcitonic state, as the emission energy of a single photon from an $N_{ex}$ state will have a red-shift of $2(N_{ex} - 1)\sum_\omega \tilde{S}_{1\omega}\hbar\omega$ (~265 meV for $N_{ex} = 20$) relative to the emission from the $N_{ex} = 1$ state. To investigate this, we perform time-resolved fluorescence upconversion photoemission spectroscopy (FLUPS) experiments. In these measurements, the photoluminescence (PL) from all NCs pumped by the Gaussian profile pump pulse is collected, and a large portion of the measured signal, and the peak of the emission, stem from emission from the large number of weakly pumped NCs at the periphery of the beam with $N_{ex} \leq 1$ (**Fig. S9**). We therefore focus our attention on the low energy tails of the emission. In the FAPbBr$_3$ NCs, we observe a strongly-redshifted contribution to the PL at short times, which increases with increasing fluence (**Fig. 5a**), as highlighted by an exponential fit to the tails of the emission at 3 ps as shown in **Fig. 5b**. At the highest pump fluences, finite PL is observed all the way to the edge of the detector, ~400 meV below the $N_{ex} = 1$ peak. This provides an independent confirmation of the large reorganization energies associated with the multiexcitonic states in the FAPbBr$_3$ NCs.

While we expect only weak coupling to lower energy optical modes in CsPbBr$_3$, stronger coupling to the higher energy optical mode ($\hbar\omega \sim 17\ meV$) have been reported, with one estimate giving $\tilde{S}_{1\omega} \sim 0.39$ to the 17 meV mode.[12] Assuming an equivalent scaling, $\tilde{S}_{N_{ex}\omega} \propto N_{ex}^2$, we could similarly expect strong red-shifted emission from the CsPbBr$_3$ NCs at high excitation densities (~250 meV for 20 excitons). However, the measurements on CsPbBr$_3$ NCs show a far weaker redshifting of the emission under the same conditions (**Fig. 5b**, **Fig. S10**) indicating no constructive interference of the lattice reorganization (i.e., $N_{ex}^2$ scaling). This may be linked to the fact that coupling to the ~17 meV mode drives symmetry lowering Jahn-Teller like tetragonal distortions of the octahedra.[27]

The nonlinear scaling of $\tilde{S}_{N_{ex}\omega}$ to low energy optical phonons implies an effective phonon-mediated attractive interaction between excitons, as the total reorganization energy associated with $N_{ex}$ overlapping excitons, $\sum_\omega \tilde{S}_{1\omega} N_{ex}^2 \hbar\omega$, is greater than that of $N_{ex}$ spatially separated excitons, $N_{ex} \sum_\omega \tilde{S}_{1\omega} \hbar\omega$. This contributes to the total binding energy between excitons, $E_B =$

$E_{B,C} + E_{B,P}$, where $E_{B,C}$ is the binding energy associated with Coloumb interaction between excitons, and $E_{B,P} = (N_{ex}^2 - N_{ex})\sum_\omega \tilde{S}_{1\omega}\hbar\omega$ is the contribution from EP-coupling. For a biexciton in FAPbBr$_3$, our extracted couplings (**Fig. 3e**) gives an $E_{B,P}(\text{FA}) \sim 14$ meV. The weak coupling to low energy optical modes in CsPbBr$_3$, combined with the lack of $N_{ex}^2$ scaling to the ~17 meV mode, implies a negligible contribution in $E_{B,P}(\text{Cs}) \sim 0$. This is consistent with the 15 meV difference in the measured biexciton binding energies in LHP NCs, where $E_B(\text{FA}) \sim 25$ meV[16] while $E_B(\text{Cs}) \sim 10$ meV.[48]

While this attractive interaction between excitons will be short lived as a result of multiexciton lifetimes (~30-50 ps),[35,36] equivalent deformation potential couplings and phonon-mediated effective attractive interactions are expected for both bare electrons and holes (**Supplementary Note 4**). Additionally, we find the magnitude of the coupling to increase with increasing temperature (**Fig. 4b**), which can enable the persistence of correlation effects at elevated temperatures.

In conclusion, femtosecond electron scattering approaches enable direct measurement of the lattice reorganization associated with interband excitation and the corresponding exciton-phonon coupling strengths in perovskite nanocrystals to be extracted. We show that deformation potential coupling of excitons to the lattice concurrently drives a change of the location of the Pb and Br ions (with a widening of the Pb-Br-Pb bond angle) on few picosecond time-scales, defining new possibilities for manipulation of their 3D crystalline structure. Our measurements and modelling indicate a coherent interference of the lattice reorganization associated with each excitation (exciton/electron/hole) in LHPs. This mediates an effective attractive interaction between excitons, electrons, and holes in LHPs, and raises the possibility of correlated charge carrier transport playing a role in their opto-electronic properties.

**Acknowledgements**
N.Y. and M. J. acknowledge support from the Swiss National Science Foundation through the Project No. 175889, the Quantum Sciences and Technology NCCR, and the Swiss National Supercomputing Centre (CSCS; project IDs s1003). M.I.B. and M.V.K. acknowledge financial support from US Air Force Office of Scientific Research (award number FA8655-21-1-7013), the Swiss Innovation Agency (Innosuisse, grant 32908.1 IP-EE), and the Swiss National Science Foundation (grant number 200021_192308, project Q-Light). The authors are grateful for the use of facilities at the Empa Electron Microscopy Center, and we acknowledge Ihor Cherniukh and Rolf Erni for performing high-resolution STEM. A.G. and N.M thank the Italian MUR for partial funding (PRIN-2017L8WW48. F.B. acknowledges support by Fondazione Cariplo (Project 2020-4382) Antonio Cervellino and the technical staff of the MS-X04SA beamline of the Swiss Light Source (Paul Scherrer Institute, CH) are acknowledged for $FAPbBr_3$ temperature-dependent X-ray measurements. B.L.C. was supported as part of the DOE 'Photonics at Thermodynamic Limits' Energy Frontier Research Center under grant DE-SC0019140. M.Y. acknowledges funding from the European Research Council (ERC) under the European Union's Horizon 2020 research and innovation programme (grant agreement No. 852751). EV and IF thank the SNSF (project 200020-184607) and the University of Geneva for financial support. The experiment was performed at SLAC MeV-UED and supported in part by the U.S. Department of Energy (DOE) Office of Science, Office of Basic Energy Sciences, SUF Division Accelerator & Detector R&D program, the LCLS Facility, and SLAC under contract Nos. DE-AC02-05CH11231 and DE-AC02-76SF00515.

**Author Contributions**
The study was devised by N.Y., V.W., A.M.L., and M.V.K., Samples were synthesized and prepared by M.B., MeV-UED experiments were performed by N.Y., B.G., B.C., M.Z., and A.M.L. supported at the beamline by M.L., M.K., A.R., X.S., S.W., and X.W., WAXTS measurements, analysis, modelling of the split-cubic structure, and diffraction simulations were performed by F. B., N. M., and A. G., FLUPS measurements were performed by I. F. and E. V., analysis of experimental data, interpretation of results, modelling of EP-coupling, and DFT calculations were preformed by N.Y., and the munscript was written by N.Y., A.M.L., and V.W., with input from all other authors.

**Additional information** Supplementary information is available for this paper at https://doi.org/xxxxxxxx. Correspondence and requests for materials should be addressed to …

**Code and data availability:** *Data will be made available on public repository upon acceptance of the manuscript.* The DebUsSy program suite is freely available at https://debyeusersystem.github.io

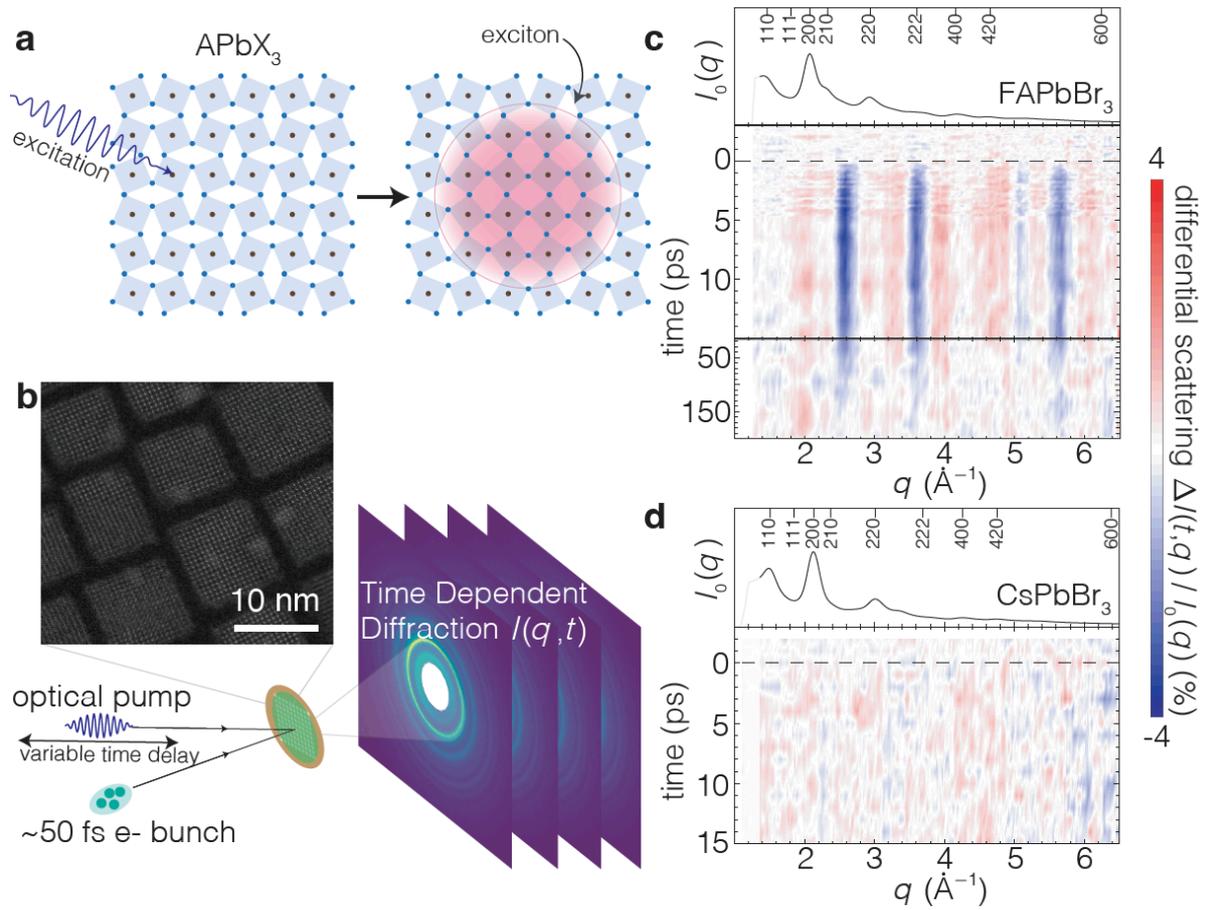

*Figure 1: Time-resolved optical-pump-electron-probe measurements of formamidinium lead bromide nanocrystals* a) illustration of a lattice reorganization of LHPs upon photoexcitation b) Schematic of the experiment including a high resolution transmission electron microscope image of $FAPbBr_3$ NCs. c) Normalised time resolved differential scattering of optically pumped $FAPbBr_3$ NCs NCs measured at 100 K with a pump fluence of 0.8 mJ/cm². A strong and ultrafast reorganization of the $FAPbBr_3$ lattice is observed upon photoexcitation. The solid black line above is the equilibrium diffraction from the NCs, and marked Bragg peaks use the hkl of the cubic phase. d) Same as panel c, for $CsPbBr_3$ NCs.

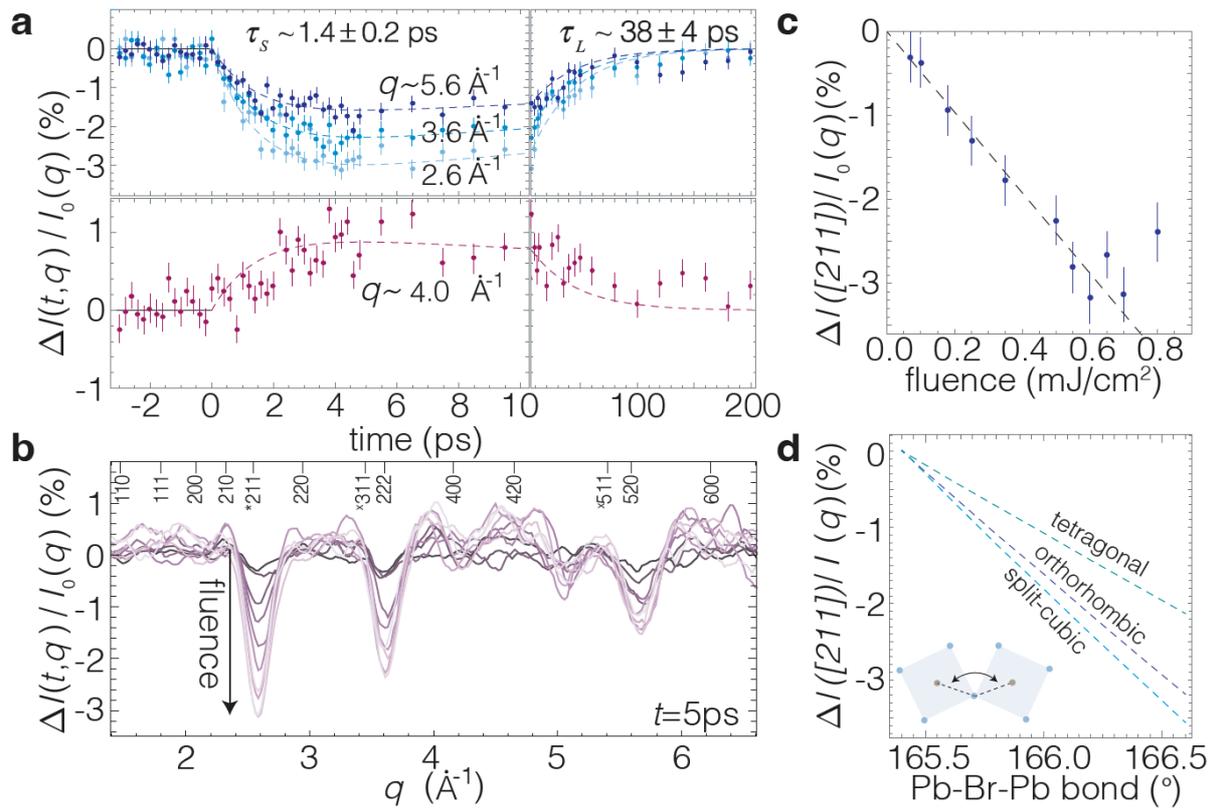

*Figure 2 Picosecond lattice reorganization of FAPbBr₃ NCs upon photoexcitation* a) Plot of the differential scattering at specific q with bi-exponential fit to the dynamics (dashed line, see **Fig. S2**) b) Differential scattering at 5 ps for fluences ranging from 0.07 mJ/cm² (darkest) to 0.7 mJ/cm² (lightest). Main Bragg peaks are marked, along with weaker reflections corresponding to peaks which are minimized in the cubic phase and sensitive to octahedral tilting (*) and Pb off-centering (ˣ). c) Fluence dependence of the strength of the photo-induced lattice reorganization as extracted from changes in the [211] peak at ~2.6 Å⁻¹ d) Simulated decrease in the intensity of the [211] reflection (~2.6 Å⁻¹) as a function of (primary-) Pb-Br-Pb bond angle for a variety of LHP structures.

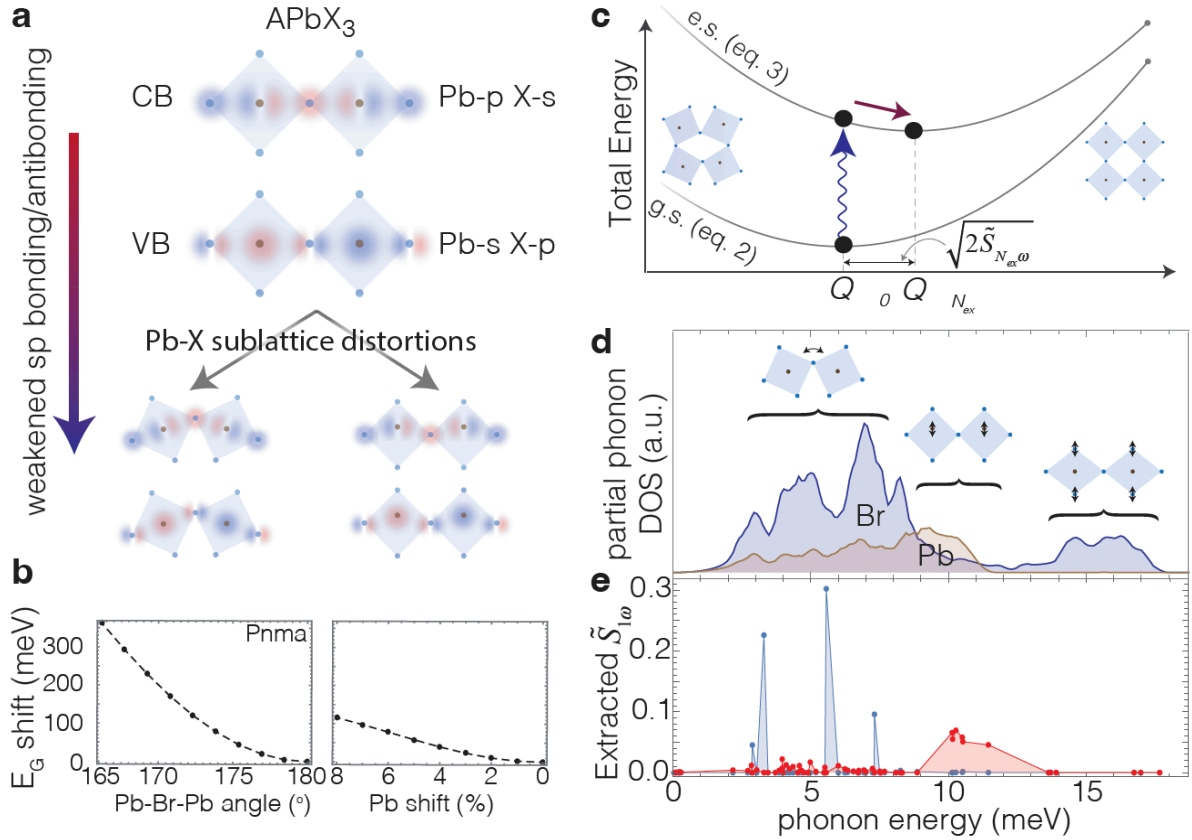

***Figure 3 Model for EP-coupling resulting from distortions of the Pb-X sublattice*** a) *Cartoon schematic of the sp bonding in the CB and VB of LHPs, along with possible symmetry lowering distortions to the Pb-X sublattice, and b) the computed shift in the energy of the bandgap for each type of distortion. Pb shift is given in percent of the nominal Pb-Br bond length. c) Model for EP-coupling to phonons driving Pb-X octahedral tilting and distortions, where the presence of excitons shifts the minimum of the total energy of the excited state towards the cubic phase. d) computed partial phonon density of states of $CsPbBr_3$, illustrations show the types of octahedral distortions driven by phonons within the specified ranges e) EP-coupling strengths resulting from the coupling of the interband excitation of a single exciton to octahedral tilting (blue points) and Pb-shifting (red points) in $FAPbBr_3$ at 100 K, extracted from the the magnitude of the measured changes in the 211 and 311 peaks respectively (Fig. 2b,d).*

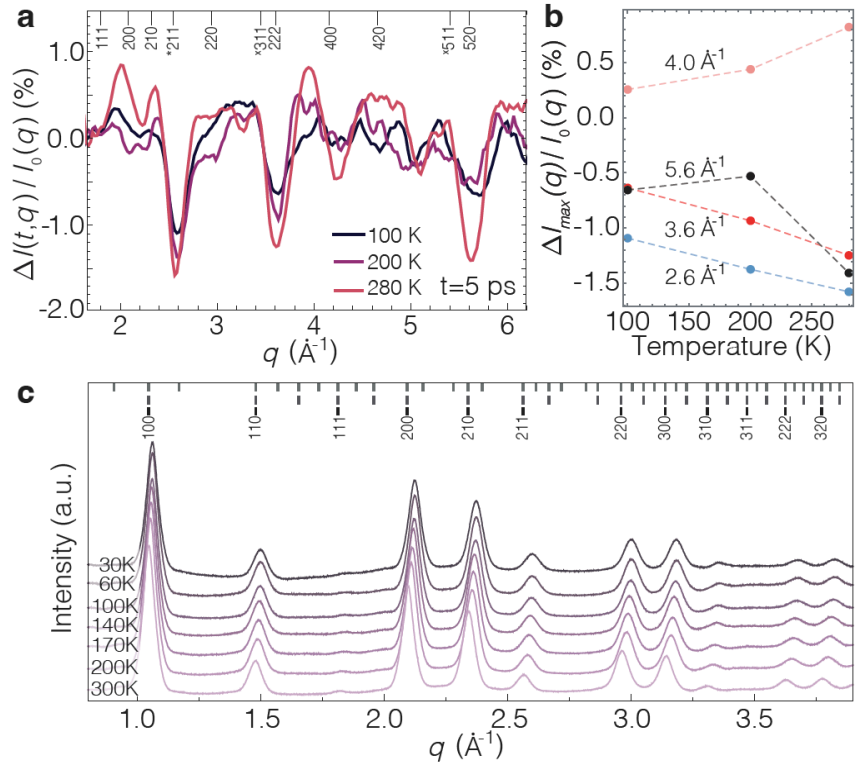

*Figure 4 Enhanced and temperature dependent EP-coupling in polymorphous FAPbBr₃ NCs a) Differential scattering measured on FAPbBr₃ NCs at 100, 200, and 280 K with a fluence of 0.5 mJ/cm², Bragg peaks are labeled as described in Fig. 2c. b) Plot of the maximum differential signal in (a) as a function of temperature at specific q values, indicating an enhancement with temperature of the photoinduced lattice reorganization. c) Temperature dependent wide angle X-ray total scattering data of FAPbBr₃ nanocrystals, collected in the 300K-30K. Ticks on the upper axis correspond to Bragg peaks of the orthorhombic (top), tetragonal (middle), and cubic (bottom) phases. The absence of characteristic superstructure peaks in the 1.6-2.0 Å⁻¹ range, as well as of other significant changes in peaks intensities among the T-dependent datasets, highlights the persistence of the polymorphic "split-cubic" structure in the whole range of temperatures explored.*

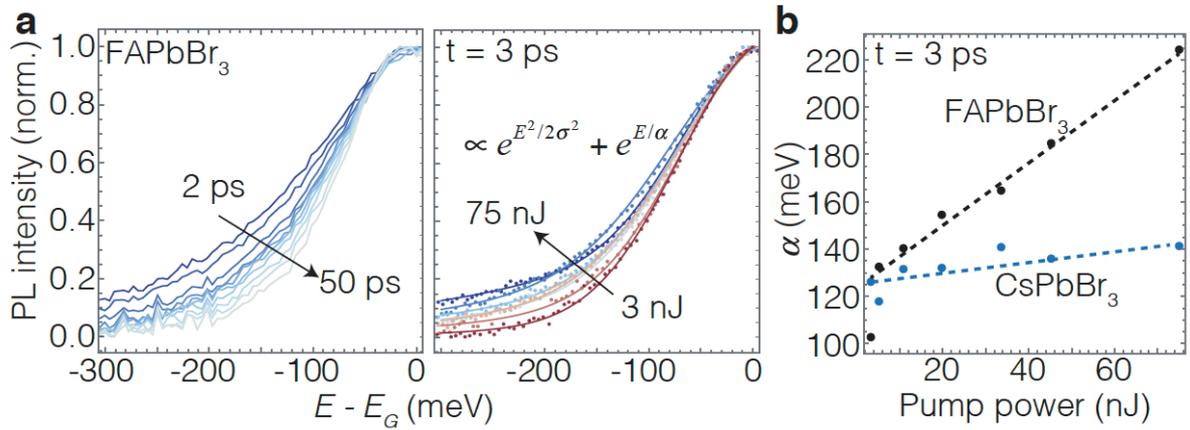

*Figure 5 Time and energy resolved fluorescence* a) Plots of the normalized emission of FAPbBr$_3$ NCs at different times for a 400nm pump pulse of 75 nJ, and at 3 ps for varying pump powers. $E_g$ is taken as the peak of the emission at large time delays. Strongly redshifted emission is observed at short time-scales in the FAPbBr$_3$ NCs. b) Plot of the energy scale of tails of the redshifted emission, $\alpha$ in (a), as a function of pump power for FAPbBr$_3$ and CsPbBr$_3$ NCs.

**Methods**

**Nanocrystal synthesis and sample preparation** Nanocrystals capped with didodecyl dimethyl ammonium bromide (DDAB) ligands were prepared using previously described procedures for $FAPbBr_3$,[30,49] and $CsPbBr_3$ NCs.[50] For time resolved electron-diffraction measurements, approximately 2 mono-layers of NCs were deposited onto 75 mesh TEM grids with amorphous carbon support (Ted Pella 01802-f). TEM grid were first fixed using anti-capillary tweezers followed by deposition of 3 μL of NC solution in mesitylene with the concentration of approx. 1 mg/mL. TEM images of the samples are given in the **Supplementary Information**.

**Optical-pump-electron-diffraction-probe measurements** Measurements were performed at the mega-electron volt ultrafast electron diffraction beamline (MeV-UED) at SLAC, the instrument is part of the LCLS User Facility. Details of the instrument have been reported previously.[51,52] 60 fs width pulses of 800 nm photons are provided by a multipass-amplified Ti:sapphire laser with a repetition rate of 360 Hz. These pulses are split, with one path generating electron beam pulses, and the other path is supplied to an optical parameteric amplifier which generates the 400 nm optical pump pulses. The generated electron bunches are accelerated to 3.7 MeV, resulting in ~10 fC pulses of ~150 fs width and 100 μm diameter at the sample. The delay between the ~500 μm optical pump pulses and the electron probe pulses were adjusted by a translational stage in the optical pumps beampath. $t = 0$ was calibrated through measurements on a bismuth thin film, while the q-scale was calibrated from single crystal diffraction on a thin film of gold. The diffracted electrons were measured via a red phosphor screen. The 2D diffraction data from the detector are azimuthally integrated into 1D diffraction profiles.

All measurements were performed on 3 identically prepared samples. Time-scans at fixed fluence were performed with a complete randomization of the pump-probe delay times of the measurement, and a new position and/or sample was choosen for each time-scan. Fluence-scans were performed with a randomization of the measured fluence at time delays of -3 ps ($I_0$) and 5 ps. The measurement time for each fluence was adjusted to obtain similar statistics for each fluence. A new position on the same sample was choosen for each fluence measurement.

**Density Functional Theory Calculations** All DFT calculations were performed using the Vienna ab initio Simulation Package (VASP),[53–56] with Projector augmented-wave (PAW) potentials [57,58] and the GGA-PBE exchange-correlation functional, with a 520 eV cutoff and a 9x9x9 gamma centered Monkhorst−Pack mesh.[59] The unit cell volume and primary tilt angle were optimized for the *Pnma* structure, the optimized structure is given in the **Supplemental Information**. Calculations of the bandgap as a function of primary *Pnma* tilt were performed on structures in which only the tilts were modified and the unit cell volume was kept constant. Calculations of the bandgap as a function of *Pnma* tilt where the unit cell volume for each tilt was optimized can be found in the **Supplementary Information**. The individual vaccum level of computed bandstructures were shifted according to minnimum of the lowest energy Cs band. For the calculation of gamma-point phonons and the phonon density of states, the atomic positions in the *Pnma* structure were optimized to a force convergence of 1 meVÅ$^{-1}$. Phonon density of states were computed using the Phonopy package[60] with a 48x48x48 gamma centered grid.

**Synchrotron Wide Angle Total Scattering (WAXTS) data collection and reduction:** WAXTS measurements on $FAPbBr_3$ nanocrystals were performed at the MS- X04SA beamline of the Swiss Light Source (Paul Scherrer Institute, Villigen, CH),[61] by drying a toluene colloidal suspension inside a 0.5 mm borosilicate glass capillary of certified composition (Hilgenberg GmbH G50). The filled capillary was fastened with a special glue inside a He-cryostream, to perform low-temperature measurements in the 300K-30K range. A beam energy of 22 keV was set and the operational wavelength ($\lambda$ = 0.563553 Å) was accurately determined using a silicon powder standard (NIST 640d, $a_0$ = 0.543123(8) nm at 22.5°C). Data were collected in the 0.4°-130° 2θ range using a single-photon counting silicon microstrip detector (MYTHEN II).[62] Background scattering from the sample holder and from the empty glass capillary were independently collected under the same experimental conditions. Angle-dependent intensities corrections were applied to the raw data to account for sample attenuation due to absorption effects; sample absorption curves were determined using an X-ray tracing method[63] and by measuring the transmitted beam from the filled capillary at room temperature, while for empty capillary the X-ray attenuation coefficient was computed using its nominal composition. Angular calibrations were applied to the zero angle of the detector and to the x, y capillary offsets, derived from the certified silicon powder standard (NIST 640d) using locally developed procedures. Background and (absorption-corrected) capillary scattering contributions were subtracted from the T-dependent sample signals.

**Time-resolved fluorescence upconversion photoemmision spectroscopy measurements**
The used setup is similar to that previously reported.[64] Excitation is provided by 100 fs at 400 nm pulses generated by doubling a portion of the output of a 1 kHz Ti:sapphire amplifier. Gate pulses of 1340 nm are produced by an optical parametric amplifier, and a CCD camera measures the upconverted spectra. Calibration was performed with secondary emissive standards.

**Methods**